\documentclass[doublecol,a4paper]{epl2}
\usepackage{graphicx}
\usepackage{ulem}
\usepackage{amssymb}
\begin{document}


\title{The Child-Langmuir law in the quantum domain}
 


\author{Debabrata Biswas and Raghwendra Kumar} 

\institute{Theoretical Physics Division,
Bhabha Atomic Research Centre,
Mumbai 400 085, INDIA}


\abstract{
It is shown using dimensional analysis that the maximum current density 
$J_{QCL}$ transported on application of a voltage $V_g$ across a gap of size $D$  
follows the relation  $J_{QCL} \sim \hbar^{3 - 2\alpha} V_g^\alpha/D^{5 - 2\alpha}$. 
The classical Child-Langmuir result is recovered at $\alpha = 3/2$ on demanding that 
the scaling law be independent of $\hbar$. For a nanogap in the deep quantum
regime, additional inputs in the form of appropriate boundary conditions and
the behaviour of the exchange-correlation potential 
show that $\alpha = 5/14$. This is verified numerically for several nanogaps.
It is also argued that in this regime, the limiting mechanism is quantum reflection 
from a downhill potential due to a sharp change in slope seen by the electron
on emerging through the barrier.
}






\maketitle

\newcommand{\be}{\begin{equation}}
\newcommand{\ee}{\end{equation}}
\newcommand{\bea}{\begin{eqnarray}}
\newcommand{\eea}{\end{eqnarray}}
\newcommand{\Tbar}{{\bar{T}}}
\newcommand{\En}{{\cal E}}
\newcommand{\Lop}{{\cal L}}
\newcommand{\DB}[1]{\marginpar{\footnotesize DB: #1}}
\newcommand{\q}{\vec{q}}
\newcommand{\kt}{\tilde{k}}
\newcommand{\Lopn}{\tilde{\Lop}}
\newcommand{\noi}{\noindent}
\newcommand{\ovn}{\bar{n}}
\newcommand{\ovx}{\bar{x}}
\newcommand{\ovE}{\bar{E}}
\newcommand{\ovV}{\bar{V}}
\newcommand{\ovU}{\bar{U}}
\newcommand{\ovJ}{\bar{J}}
\newcommand{\calE}{{\cal E}}



\section{Introduction}
\label{sec:intro}

The Child-Langmuir (CL) law \cite{CL} is the cornerstone of classical space charge limited
flows. It determines the {\it maximum current} ($J_{CL}$) that can flow across
a gap of size $D$ on application of an electric field $-V_g/D$, given the existence of
mono-energetic electrons at the cathode \cite{other_interpretations,pop3}.
The limiting mechanism is fairly straightforward: the mutual repulsion 
between the electrons gives rise to a 
barrier in the net electrostatic potential and an energy spread in the initially mono-energetic 
beam of electrons. Beyond the limiting current, the excess electrons have energy slightly lower 
than the barrier height and are thus reflected back to the cathode. This picture 
holds for electrons with zero energy as well (relative to the cathode potential)
where a more common, albeit equivalent, criterion of the limiting mechanism 
is the total cancellation of the applied field by the
space charge field at the cathode and can thus be arrived at from the Poisson equation
of electrostatics.

The quantum treatment for studying steady state charge transport across 
a gap is based on a self-consistent solution of the Schr\"{o}dinger
and Poisson equations \cite{lau,ang2003}. As in the classical case,
there exists a maximum current in this formalism beyond which
a solution to the coupled set of equations ceases to exist.
The exact limiting mechanism is however largely uninvestigated
with the studies largely centred around scaling laws.

The purpose of this letter is to (i) provide a unified scaling
law across all regimes whenever a power law behaviour holds,
(ii) probe the deep quantum regime using additional inputs
such as the boundary conditions and the behaviour of the exchange-correlation
potential and (iii)  investigate the limiting
mechanism in the quantum domain.

The model that we study pre-supposes
the existence of free mono-energetic electrons at the cathode, whether 
due to thermionic, field or photo emission.  As in case of the classical
Child-Langmuir law, the quantum framework merely seeks to predict the 
maximum steady current that can be
transported on application of a field. Importantly, the classical or quantum Child-Langmuir law 
does not in itself address the question of emission. Together with emission
laws  however, the
limiting Child-Langmuir current plays a role in understanding
current-voltage characteristics.

Quantum effects become important at low applied
voltages or small gap size and give rise to voltage scaling 
other than the classical three-halves law ($V_g^{3/2}$), a
subject of considerable interest due to the current focus on
nanostructured materials. 
In the deep quantum limit, the small phase volume permits 
few electrons so that the
Hartree potential is inconsequential and the exchange correlation
potential dominates. 
We shall focus on this regime in this communication
and try to understand the limiting mechanism and 
explore the scaling behaviour with respect to the applied potential.

The quantum  phenomenon of reflection, tunneling and 
the fermionic nature of electrons are best incorporated within the 
Kohn-Sham density functional
theory \cite{KS} with an effective potential, $V_{eff}$, that includes a parametrized form
of the exchange-correlation potential (such as due to Perdew and Zunger \cite{PZ})
obtained within the local density
approximation (LDA). Since, the emission mechanism is not considered here,
limitations of LDA, such as the fact that it does not correctly
reproduce the image potential, is not of much consequence.

The basic inputs for implementing Kohn-Sham theory 
to study transport across nanogaps are twofold. The first concerns the 
boundary conditions for the Hartree potential
at the gap boundaries. Since the applied voltage difference 
is proportional to the difference in chemical potential between the 
injection and collector planes, the exchange-correlation potential
at the two ends play a significant role in determining the boundary
conditions in the deep quantum regime and its neglect can lead to
erroneous results \cite{comment}. The second input
lies in the determination of the boundary conditions for the
wavefunction and is related to assumptions for the potential
beyond the gap for purposes of matching the wavefunction. 
The simplest of these is to assume that the effective potential 
assumes a  constant value (abruptly) on either side of the gap.
This approach is however known to give misleading results 
when the exchange-correlation
potential   is neglected altogether \cite{db2012}. 
First, an abrupt change in potential for the Schr\"{o}dinger
equation  can result in considerable quantum reflection 
at low applied voltages or small gap size.
Further, the coupling of the Schr\"{o}dinger and Poisson equations makes the
boundary conditions for the wavefunction and Hartree potential
inconsistent \cite{db2012}.
The problem can be addressed by simply assuming that the transported charged
particles are not reflected back into the gap and that the
effective potential varies smoothly across the interface.
An approximate transmitted wavefunction is then readily  
available within the semiclassical approximation, which can
then be matched with the wavefunction in the nanogap at
the collection plane.

\section{Formalism}
\label{sec:formalism}

We shall first familiarize ourselves with the equations to be solved
self-consistently and subsequently arrive at the boundary conditions
that they need to satisfy. The effective potential energy, $V_{eff}$,
in the Schr\"{o}dinger equation

\be
-d^2 \psi/dx^2 + V_{eff} \psi = E \psi \label{eq:Schrod}
\ee

\noi
takes the form $V_{eff} = -eV + V_{xc} \times E_H$, where 
$E_H = e^2/(4\pi\epsilon_0 a_0)$ is the 
Hartree energy and $a_0$ the Bohr radius. The exchange correlation
potential within LDA takes the form 
$V_{xc} = \epsilon_{xc} - (r_s/3) d\epsilon_{xc}/dr_s$, where
$r_s = [3/(4\pi n)]^{1/3}$ is 
the Wigner-Seitz radius, $n$ the electron 
number density and $e$ the magnitude of the electronic charge. In the above,
$V$ is the Hartree potential satisfying the Poisson equation

\be
d^2 V/dx^2 = e n(x)/\epsilon_0 =  e |\psi(x)|^2/\epsilon_0 \label{eq:Poi}
\ee

\noi
with boundary conditions that we shall shortly specify.
The exchange correlation energy 
density,  $\epsilon_{xc}$ is expressed
as $\epsilon_{xc} = \epsilon_x + \epsilon_c$ where 
$\epsilon_x = -(3/2\pi)^{2/3} (3/4r_s)$
is the exchange contribution for a uniform electron gas while
$\epsilon_c$, the correlation contribution, is represented by a 
parametrized form of the random phase approximation result.
Unless otherwise specified, we shall restrict ourselves to the
Perdew-Zunger \cite{PZ} parametrization.

We first note that the applied voltage difference, $V_g = -(\mu_C - \mu_E)/e$ 
where $\mu_C$ and $\mu_E$ refer respectively to the chemical potential at the collector and
injection planes. For convenience, we consider the reference as
$\mu_E = -eV(0) + V_{xc}(0)\times E_H = 0$ so that $E = 0$ refers to injection from
the Fermi level. Thus
$V(0) = V_{xc}(0) \times E_H/e$. It follows that the chemical potential at the
collector is $-eV(D) + V_{xc}(D) \times E_H = -eV_g$. 
Thus $V(D) = V_g + V_{xc}(D) \times E_H/e$.
In writing the above, we have implicitly assumed continuity of
the chemical potential at the interfaces under steady-state conditions.  

The coupled Schr\"{o}dinger-Poisson system with the above boundary conditions for $V$
are to be solved self-consistently under the assumption that a current density $J$ is transported
across the collection plane. This accounts for one of the two boundary
conditions required to solve the Schr\"{o}dinger equation. The other is 
derived under an approximation.

The simplest of these is the assumption that the effective potential is constant
beyond the collection plane. As discussed in the introduction however, this can
lead to erroneous results, mainly because a self-consistent solution requires
$V_{eff}'(D) = 0$. If the exchange-correlation contribution is small, this
implies $V'(D) = 0$, a condition that is found to be violated
except in regimes where a classical description provides an adequate
solution.

At the next level, assuming that
there is no reflection back into the nanogap and that the effective
potential varies sufficiently smoothly at the interface, the transmitted
solution can be expressed as a first order WKB wavefunction

\be
\psi_{trans} \simeq \frac{C}{\sqrt{p(x)}} e^{ \frac{i}{\hbar} \int^x p(x') dx'} \label{eq:wkb}
\ee

\noi
where $p(x) = \sqrt{2m(E - V_{eff}(x))}$ is the classical momentum. The constant $C$ 
can be fixed in terms of the known quantities $J$, $E$ and $V_{eff}(D)$.
The approximate transmitted wavefunction, $\psi_{trans}$ can then be used to match
the gap wavefunction $\psi$ and its derivative at the boundary.

The coupled system of equations can be solved by first expressing the 
complex wavefunction 

\be
 \psi = \sqrt{n_0} q(x) e^{i\theta(x)} \label{eq:psi}
\ee

\noi
in terms of a real amplitude 
$q(x)$, phase $\theta(x)$ and the characteristic density 
$n_0 = 2\epsilon_0 V_g/3eD^2$. Using dimensionless normalized variables
$\ovx = x/D$, $\phi = V/V_g$, $\lambda = D/\lambda_0$ where
the electron de Broglie wavelength $\lambda_0 =  \sqrt{\hbar^2/2meV_g}$,
$\epsilon = E/eV_g$, $\ovJ = J/J_{CL}$
the Schr\"{o}dinger and Poisson equations can be expressed respectively 
as \cite{ang2003}

\be
\frac{d^2q}{d\ovx^2} + \lambda^2 [\epsilon + \phi  - \phi_{xc} - 
\frac{4}{9} \frac{\ovJ^2}{q^4} ]q = 0 ~~~~~{\rm and} \label{eq:Sch}
\ee

\be
\frac{d^2\phi}{d\ovx^2} =  \frac{2}{3} q^2 \label{eq:Poi1}
\ee

\noi
where $\phi_{xc} = V_{xc}/\phi_g$, $\phi_g = eV_g/E_H$, 
$V_{xc} = \epsilon_{xc} - (r_s/3) d\epsilon_{xc}/dr_s$,

\bea
\epsilon_{x} & = & -\frac{3}{4} (\frac{3}{2\pi})^{2/3}\frac{1}{r_s}, \\
\epsilon_{c} & = & \frac{\gamma}{1 + \beta_1 \sqrt{r_s} + \beta_2 r_s} \label{eq:PZ}
\eea

\noi
with $r_s = [3\lambda/2\phi_g q(\ovx)]^{2/3}$, $\gamma = -0.1423$, $\beta_1 = 1.0529$
and $\beta_2 = 0.3334$. The parametrized form for the correlation energy density
in Eq.~\ref{eq:PZ} is due to Perdew and Zunger \cite{PZ} with the 
parameters obtained by fitting to the
random phase approximation results.

With the above scaling, the boundary conditions for the scaled
Hartree potential (Eq.~\ref{eq:Poi1}) are 

\bea
\phi(0) & = & \phi_{xc}(0) \label{eq:phi@0} \\
\phi(1) & = & 1 + \phi_{xc}(1).  \label{eq:phi@1}
\eea

The boundary conditions for $q(\ovx)$ can be derived under the assumption that
the transmitted wavefunction is given by Eq.~\ref{eq:wkb} and carries a 
current density $J$. Using Eq.~\ref{eq:wkb} and 
$J = e i\hbar ( \psi^* \psi^{'} - \psi \psi^{*'}) / {2m}$, it follows
that $|C|^2  = Jm/e$. Without any loss of generality, we shall assume 
that $C$ is real. Thus $C = \sqrt{Jm/e}$. On equating $\psi$ (Eq.~\ref{eq:psi})
and $\psi_{trans}$ (Eq.~\ref{eq:wkb}) at $x = D$ ($\ovx = 1$), we
have 

\bea
q(1) & = & \frac{\sqrt{\frac{2\ovJ}{3}}}{[\phi(1) + \epsilon - \phi_{xc}(1)]^{1/4}}  \\
     & = & \frac{\sqrt{\frac{2\ovJ}{3}}}{[1 + \epsilon]^{1/4}} \label{eq:q1}
\eea

\noi
where Eq.~\ref{eq:q1} follows on using the boundary condition $\phi(1) = 1 + \phi_{xc}(1)$.
In arriving at Eq.~\ref{eq:q1}, we have equated $\theta(1) = (1/\hbar)\int_{x_r}^x p(x') dx'$,
in effect fixing the value of the reference point  $x_r$.

The boundary condition for $dq/d\ovx$ can be similarly derived by equating the
derivatives of $\psi$ and $\psi_{trans}$. However, since $\phi_{xc}$ depends 
on $q(\ovx)$, the value of $q'(1)$ must be computed from the equation

\bea
q'(1) = - \frac{1}{4} \sqrt{\frac{2\ovJ}{3}} 
\frac{[\phi'(1) - V_{xc}'(1)/\phi_g]}{[\phi(1) + \epsilon - \phi_{xc}(1) ]^{5/4}}
\eea

\noi
where the primes indicate differentiation with respect to $\ovx$.
Since 

\be
\frac{dV_{xc}}{d\ovx}|_{\ovx=1} = \frac{dV_{xc}}{dr_s} |_{r_s(1)} \times \frac{dr_s}{d\ovx}|_{\ovx=1}
\ee
\noi
and 

\be
 \frac{dr_s}{d\ovx}|_{\ovx=1} = - (\frac{3\lambda}{2\phi_g})^{2/3} \frac{2}{3} \frac{1}{q^{5/3}(1)} q'(1),
\ee
\noi
it follows that

\be
q'(1) =  - \frac{\frac{1}{4} \sqrt{\frac{2\ovJ}{3}} \phi'(1)}{[1 + \epsilon]^{5/4}} \frac{1}{\zeta} \label{eq:q1p}
\ee

\noi
where

\be
\zeta = 1 + \frac{\frac{1}{6}\sqrt{\frac{2\ovJ}{3}} (3\lambda/2)^{2/3} (dV_{xc}/dr_s)|_{r_s(1)}}{[1 +  \epsilon]^{5/4}}
\ee

\noi
As $V_{xc} \rightarrow 0$, $\zeta \rightarrow 1$ so that $q'(1)$ takes the value
predicted on ignoring exchange-correlation effects altogether \cite{db2012}.


\section{Voltage Scaling: Dimensional Analysis}
\label{sec:scaling_dimensional}

\begin{figure}[tbh]
\begin{center}
\includegraphics[width=5.5cm,angle=270]{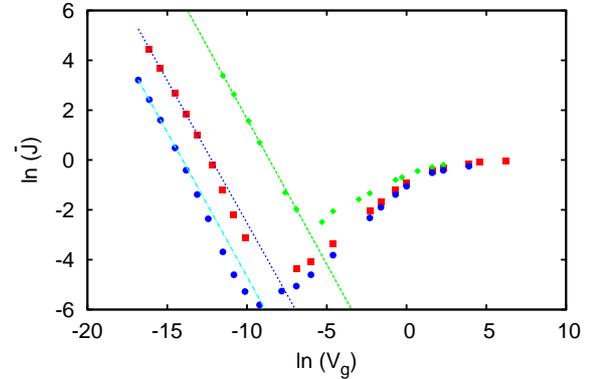}
\end{center}
\caption{The maximum scaled current density $\ovJ$ vs $V_g$ for
three different nanogaps (i) $D=10$nm - top set of points marked as diamonds 
(ii) $D = 30$nm - solid squares in the middle (iii) $D = 50$nm - the bottom set 
marked by solid circles. The best fitting straight line in the deep quantum 
regime is also shown. In all three cases, the slope is in the range -1.16 to -1.14.
Note that the voltage at which the deep quantum regime sets in is smaller
for larger gaps.}
\label{fig:Jbar_vs_Vg}
\end{figure}

As the Schr\"{o}dinger equation dictates the deep quantum regime, 
the scaling with applied voltage should differ from the 
Poisson equation induced $V_g^{3/2}$ scaling. 
The existence of a power law behaviour in the quantum regime cannot
be assumed outright. However, numerical results clearly show a power
law behaviour in the deep quantum regime (see Fig.~\ref{fig:Jbar_vs_Vg}).

Expressing the scaling in terms of dimensionless quantities 
as $\ovJ \sim (V_g/V_s)^\beta$ where $\ovJ = J/J_{CL}$
and the voltage scale $V_s = \hbar^2/(2meD^2)$, it follows that

\be
J \sim \frac{ \epsilon_0 e^{\beta + 1/2} \hbar^{-2\beta}}{ m^{1/2 - \beta}}
\frac{V_g^{\beta + 3/2}}{D^{2-2\beta}}.
\ee

\noi
Writing $\beta + 3/2 = \alpha$, it follows that 

\be
J \sim \epsilon_0 \frac{\hbar^{3-2\alpha}e^{\alpha-1}}{m^{2-\alpha}}\frac{V_g^\alpha}{D^{5-2\alpha}} 
\label{eq:scaling_expt}
\ee

\noi
giving the scaling $J \sim V_g^\alpha/D^{5-2\alpha}$ as shown in \cite{db2012}.

The scaling law above is based on numerical observations. It can however
be derived from purely dimensional analysis. The relevant equations are
the Schrodinger (Eq.\ref{eq:Schrod}) and Poisson equations (Eq.~\ref{eq:Poi})
together with the current equation

\be
J = \frac{e\hbar }{2im} [\psi^* \nabla{\psi} - \psi \nabla{\psi^*}].
\ee 

\noi
Redefining $\tilde{\psi} = \psi/\sqrt{\epsilon_0}$, it follows that 
the Schrodinger and Poisson equations are free of $\epsilon_0$
while the current density takes the form

\be
J = \epsilon_0 \frac{e \hbar}{2im} [\tilde{\psi}^* \nabla{\tilde{\psi}^*} 
- \tilde{\psi}^* \nabla{\tilde{\psi}} ].
\ee 

\noi
Thus, $\tilde{\psi}$ no longer depends on $\epsilon_0$ so that
$J \sim \epsilon_0$. However $\tilde{\psi}$ continues to 
depend on $e$, $\hbar$ and $m$ and hence their dependence on
$J$ need not be linear. We are now in a position to implement
standard dimensional analysis to ascertain the scaling law.

Since, the critical current can in general depend on the quantities $V_g$,
$D$, $\epsilon_0$, $\hbar$, $m$ and $e$, we demand

\be
J^{-1} V_g^{\alpha} D^{\delta_1} \hbar^{\delta_2} \epsilon_0 m^{\delta_3} e^{\delta_4} = C_0
\ee

\noi
where $C_0$ is a dimensionless constant. On expressing all the quantities
in terms of mass, length, time and current ($M,L,T,A$), and equating
powers of these to zero, we obtain four equations

\bea
\delta_4  - \alpha & = & -1 \\
2\delta_2 + 2\alpha + \delta_1 & = & 1 \\
\delta_4 - \delta_2 - 3\alpha & = & -4 \\
\delta_2 + \delta_3 + \alpha & = & 1 
\eea

\noi
in terms of five unknown quantities. Expressed in terms of $\alpha$, the
exponent of $V_g$, we obtain $\delta_1 = 2\alpha$ - 5, $\delta_2 = 3 - 2\alpha$,
$\delta_3 = \alpha - 2$ and $\delta_4 = \alpha - 1$. Thus, standard dimensional
analysis gives us Eq.~\ref{eq:scaling_expt} confirming that any 
power law behaviour must follow this relation.

Note that the Child-Langmuir law follows automatically
on demanding that the exponent of $\hbar$ be zero. Thus $\alpha = 3/2$
so that $J \sim V_g^{3/2}/D^2$ \cite{so_called,ang2004}.

\section{Scaling in the deep quantum regime}
\label{sec:scaling}

While dimensional analysis leads us to the general scaling law 
$J \sim \hbar^{3 - 2\alpha} V_g^\alpha/D^{5 - 2\alpha}$, the value of $\alpha$
observed in Fig.~\ref{fig:Jbar_vs_Vg} can be understood by the following
analysis that involves additional inputs. We first note
from Eq.~(\ref{eq:q1}) that  

\be 
\ovJ \sim q^2(1).
\ee

\noi
To determine the dependence of $q^2(1)$ on $V_g$, we shall investigate
the quantity

\be
\frac{d}{dV_g}q^2(1) = \frac{d}{dV}q^2(\ovx)|_{\ovx=1}
\ee

\noi
where the equality follows on using the alternate but equivalent 
boundary conditions for the electrostatic potential:

\bea
V(\ovx = 1) & = & V_g  \\
V(\ovx = 0) &  = & \frac{E_H}{e} (V_{xc}(0) - V_{xc}(1))
\eea 

\noi
with the injection energy redefined equivalently as $E = V_{xc}(1) E_H$. 
Note that this does not result in any change in the expression for $q(1)$ or
$q'(1)$ since $\phi(1) + \epsilon - \phi_{xc}(1)$ remains invariant (equals
unity) for both sets of boundary condition and injection energy.

Further, since 

\be
\frac{d}{dV}q^2(\ovx=1) = (\frac{d}{d\ovx}q^2 / \frac{d}{d\ovx}V)_{|_{\ovx=1}}
\ee

\noi
we have 

\bea
\frac{d}{dV_g}q^2(1) & = & 2 q(1) q'(1) / V'(1) = \frac{2 q(1) q'(1)}{V_g \phi'(1)}\\
& \sim & \frac{\ovJ}{V_g \zeta}
\eea

\noi
We now note that in the quantum regime $\zeta >> 1$ so that
$\zeta \sim \lambda^{2/3} \sqrt{\ovJ} (dV_{xc}/dr_s)|_{r_s(1)}$. As
$\lambda \sim V_g^{1/2}$, it follows that 

\be
\frac{d}{dV_g}q^2(1) \sim \frac{\sqrt{\ovJ}}{V_g^{4/3} (\frac{dV_{xc}}{dr_s})_{|_{r_s(1)}} }.
\ee

\noi
Note that the electron density is small at $\ovx = 1$ compared to 
$\ovx = 0$ and is negligible at low applied voltages. Thus $r_s(1)$ 
is large in quantum regime. It can be verified that for the Perdew-Zunger
exchange-correlation potential,

\bea
\frac{dV_{xc}}{dr_s}|_{r_s(1)} & \sim & \frac{1}{r_s^2(1)} \sim n^{2/3}(1) \sim J^{2/3} \\
& = & (\ovJ J_{CL})^{2/3} \sim \ovJ^{2/3} V_g.
\eea

\noi
Thus,

\be
\frac{d}{dV_g}q^2(1) \sim \frac{1}{V_g^{7/3} \ovJ^{1/6}}
\ee

\noi
Assuming a power law behaviour $\ovJ \sim V_g^{\alpha'}$, the above equation 
translates as

\be
V_g^{\alpha' - 1 + \alpha'/6} \sim V_g^{-7/3}.
\ee

\noi
On equating the powers, we have $7\alpha'/6 = - 4/3$, or

\be
\alpha' = -8/7.
\ee

\noi
Thus, $\ovJ \simeq V_g^{-8/7}$ or $J \sim V_g^{5/14}$ so that $\alpha = 5/14$.
The value of $\alpha' = -8/7 \simeq -1.143$ agrees very well with the numerical results 
presented in Fig.~\ref{fig:Jbar_vs_Vg} where the exponent $\alpha'$ is
found to lie between -1.16 and -1.14 in all three cases.

\section{The Limiting Mechanism}
\label{sec:mechanism}

\begin{figure}[tbh]
\begin{center}
\includegraphics[width=5.5cm,angle=270]{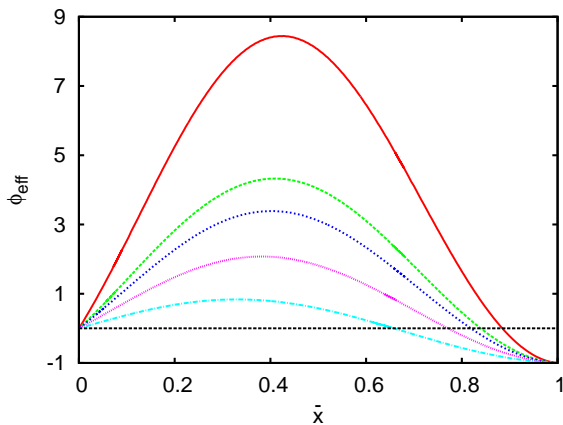}
\end{center}
\caption{Typical effective scaled potential curves for $D = 30$nm
and $V_g = 2 \times 10^{-7}$V in order of increasing current density from
bottom to top. The horizontal dashed line is $\phi_{eff} = 0$.
The values of $\ovJ$ are 20,30,35,37 and 39.85 (topmost).
The critical scaled current density is around 39.9.
}
\label{fig:potential}
\end{figure}

In the classical regime, the Schr\"{o}dinger-Poisson system decouples
and the Poisson equation alone dictates the limiting behaviour.
In the deep quantum regime of low applied voltages in nanogaps, 
numerical investigations show that $\phi(\ovx)$ is
linear and follows the applied voltage curve. Thus,
the Schrodinger equation 
alone decides the existence of a physically acceptable solution. 
We shall briefly consider the limiting mechanism for the model considered
in this communication.

In the steady state, an acceptable solution is a smooth, spatially non-oscillatory
behaviour of the effective potential.  
At very low applied voltages in a nanogap, there are too few electrons to
alter the electrostatic potential curve from the linear applied 
potential significantly. Superimposed on this however is the
repulsive exchange-correlation potential which decides the effective
potential in the Schrodinger equation. Typically, $\phi_{xc}$ is 
small at large $r_s$ and falls below the electrostatic potential
close to the right boundary. Thus the attractive potential dominates
near the anode. Near the cathode however, $r_s$ is small and 
$\phi_{xc}$ dominates over $\phi$. Typically, an acceptable limiting effective
potential in the deep quantum regime has a barrier towards the cathode, passes 
through zero and becomes attractive closer to the anode.
For $D=30$nm and $V_g = 2 \times 10^{-7}$V, the change in scaled effective potential
with current density is shown in Fig.~\ref{fig:potential}. The topmost potential curve
is close to the critical current density.
Note that such a solution ceases to exist beyond criticality.

Two observations can be made immediately from the change in potential curves
as criticality is approached. First, the barrier height and width increase
as the current density is increased. Second, in the region
where the electron is free, from the point where it tunnels through the
barrier to the collector plate at $\ovx = 1$, the change in slope increases
as criticality is approached (see Fig.~\ref{fig:potential}).
  
The increase in barrier height and width however cannot be 
the limiting mechanism for the model considered since even a tiny transmission
coefficient cannot violate any of the boundary conditions, in particular that
of an outgoing transmitted wave carrying a current density $J$ to the collector.
It of course requires a large number of electrons to be present at $\ovx = 0$, 
due, for example, to photo-emission.

\begin{figure}[tbh]
\begin{center}
\includegraphics[width=5.5cm,angle=270]{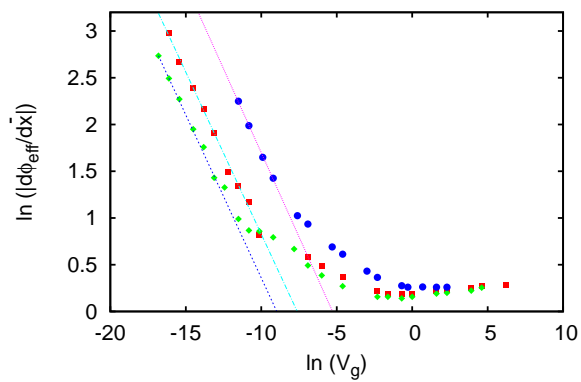}
\end{center}
\caption{The maximum slope of the effective potential in the free 
region vs $V_g$ for three different nanogaps (i) $D=10$nm - top set of points marked as circles 
(ii) $D = 30$nm - solid squares in the middle (iii) $D = 50$nm - the bottom set 
marked by diamonds. The best fitting straight line in the deep quantum 
regime is also shown. In all three cases, the slope is in the range 0.33-0.34.
The maximum slope is measured in the region where the
electron is free. In the deep quantum regime, the maximum slope  occurs at
the point where $\phi_{eff}$ becomes zero while in the classical regime, the
maximum slope shifts to $\ovx = 1$.
 }
\label{fig:slope_vs_Vg}
\end{figure} 

A sharp change in slope of the effective potential in the downhill region 
is however known to cause quantum reflection \cite{reflection,atomic_reflectometry}
just as a sudden change in refractive index leads to optical reflection. Such a 
phenomenon does occur in the approach to criticality as seen in Fig.~\ref{fig:potential}.
In order to test the hypothesis, we study the 
dependence of maximum slope on the applied voltage for different nanogaps in the free region.
Fig.~\ref{fig:slope_vs_Vg} shows a universal behaviour in the
quantum regime of low applied voltages. The dependence of maximum slope
on the applied voltage is clearly identical in all three cases
with $|d\phi_{eff}/d\ovx| \sim V_g^{\delta}$ with $\delta \simeq -0.33$ \cite{classical}.
Since the slope $d\phi_{eff}/d\ovx \rightarrow 0$ as $\ovx \rightarrow 1$,
the onset of criticality appears to be linked to the change in slope
seen by the electron as the electron emerges from the barrier.
Note that a crucial input in deriving the boundary conditions is the
assumption that an outgoing wave carries a current density $J$.
This breaks down in the presence of quantum reflection from a 
downhill potential and is possibly the limiting mechanism \cite{criterion}.

\section{Summary and Conclusions}
\label{sec:conclusions}

We have considered the problem of maximum transmitted current on application
of a bias, assuming that a sea of electrons exist at the 
end having lower potential, for example due to photo-emission. The quantum model
considered is 1-dimensional and includes exchange-correlation effects.
The applied bias is carefully fixed in terms of the difference
in chemical potential at the two ends and the boundary conditions for the
wavefunction amplitude are derived using a WKB expansion at the collector plane
to account for a consistent solution of the Poisson-Schr\"{o}dinger system.

It was shown using standard dimensional analysis that the scaling law
applicable across all regimes is  
$J_{QCL} \sim \hbar^{3 - 2\alpha} V_g^\alpha/D^{5 - 2\alpha}$. The classical
Child-Langmuir law follows on demanding that the exponent of $\hbar$
be zero. Thus $\alpha = 3/2$ and  $J_{QCL} \sim V_g^{3/2}/D^2$.

It was found numerically that for the Perdew-Zunger exchange-correlation potential: (a) the
maximum transmitted current falls below the classical prediction till the
quantum regime is reached (b) the applied voltage signalling the onset of the 
quantum regime decreases as the gap size is increased (c) in the deep quantum
regime, the maximum transmitted current density is higher than the classical
prediction and the region is also marked by a universal behaviour in the scaling
of the current density with applied voltage. The voltage scaling 
was shown analytically to follow $J \sim V_g^\alpha$  with exponent 
$\alpha = 5/14$. The exponent was found to be close to the 
numerically observed values.

We conclude this communication with a few remarks: (i) the results presented here
are directly applicable experimentally only when electrons are made available 
at the emitter end by an alternate mechanism such as photo-emission. In case of
field emission, these results can only provide an upper bound to the transmitted
current density (ii) the deep quantum regime depends sensitively on the exchange-correlation
potential and predictions of higher transmitted current and universal behaviour
may vary with a different parametrizations of $V_{xc}$. It may be hoped that careful
experimental observations will help decide the appropriate form of 
of correlation energy applicable for electron transport in nanogaps.



\begin{thebibliography}{99}
\bibitem{CL}C.~D.~Child, Phys. Rev. Ser. 1 {\bf 32}, 492 (1911); 
I.~Langmuir, Phys. Rev. {\bf 2}, 450 (1913).
\bibitem{other_interpretations} The Child-Langmuir Law for zero injection energy can be 
generalized in several ways for non-zero injection energy relative to the cathode
potential energy \cite{pop3}. For zero-injection energy, all coincide.
\bibitem{pop3} R.~R.~Puri, D.~Biswas and R.~Kumar, Phys. Plasmas {11}, 1178 (2004).
\bibitem{lau} Y.~Y.~Lau, D.~Chernin, D.~G.~Colombant and P.-T.~Ho, Phys. Rev. Lett. 66, 1446 (1991).
\bibitem{ang2003} L.~K.~Ang, T.~J.~T.~Kwan, and Y.~Y.~Lau, Phys. Rev. Lett. {\bf 91}, 208303
(2003).
\bibitem{KS} W.~Kohn and L.~J.~Sham, Phys. Rev. {\bf 140} A1133 (1965).
\bibitem{PZ} J.~P.~Perdew and A.~Zunger, Phys. Rev. B {\bf 23}, 5048 (1981) 
\bibitem{comment} D.~Biswas, Phys. Rev. Lett. {\bf 109}, 219801 (2012).
\bibitem{db2012} D.~Biswas and R.~Kumar, Eur. Phys. J. B {\bf 85} 189 (2012). 
\bibitem{so_called} The purported dimensional argument presented in \cite{ang2004}
and reiterated in \cite{ang2012} predicts a single value for the exponent $\alpha$
within the Schr\"odinger-Poisson system 
and hence does not recover the classical value of $\alpha$ from the quantum formalism. 
Numerical results however indicate a smooth transition to the classical result.
\bibitem{ang2004} L.~K.~Ang, Y.~Y.~Lau and T.~J.~T.~Kwan, IEEE Trans. Plasma Sci. {\bf 32}, 410 (2004).
\bibitem{ang2012} L.~.K.~Ang,  Phys. Rev. Lett. {\bf 109}, 219802 (2012). 
\bibitem{reflection} P.~L.~Garrido, S.~Goldstein, J.~Lukkarinen and R.~Tumulka, 
Am. J. Phys. {\bf 79}, 1218 (2011).
\bibitem{atomic_reflectometry} B.~Segev, R.~C\^{o}te and M.~G.~Raizen, Phys Rev A{\bf 56}, R3350 (1997).
\bibitem{classical} Note that classically at criticality, $\phi(\ovx) = \ovx^{4/3}$. Thus
the points converge at $d\phi_{eff}/d\ovx = 4/3$ at higher voltages
where the classical description is applicable and exchange correlation effects can be neglected.

\bibitem{criterion} While a criterion for the onset of quantum reflection in the 
downhill region has been offered before (see e.g. \cite{atomic_reflectometry}) 
and is identical to the criterion for the breakdown of WKB approximation, it is not 
our intention here to test this or offer a new criterion. 

\end{thebibliography}
\end{document}